\documentclass[aps,pra,reprint,a4paper,superscriptaddress,amsmath,floatfix,longbibliography]{revtex4-1}
\usepackage[dvips]{graphicx}

\DeclareMathOperator{\diag}{diag}
\newcommand{\fpp}[2]{\frac{\partial #1}{\partial #2}}

\newcommand{\vniia}{Dukhov Research Institute of Automatics (VNIIA), Moscow, 127055 Russia}
\newcommand{\Bilbao}{Bilbao1,*Bilbao2,*Bilbao,*BilbaoBZ,*BilbaoBZ2}

\begin{document}

\title{Strain-induced pseudomagnetic field in Dirac semimetal borophene}
\author{A.\,D.\,Zabolotskiy}
\email{zabolotskiy@vniia.ru}
\affiliation{\vniia}
\author{Yu.\,E.\,Lozovik}
\email{lozovik@isan.troitsk.ru}
\affiliation{\vniia}
\affiliation{Institute for Spectroscopy, Russian Academy of Sciences, Troitsk, Moscow, 142190 Russia}
\affiliation{Moscow Institute of Physics and Technology, Dolgoprudny, Moscow region, 141700 Russia}
\affiliation{National Research Nuclear University Moscow Engineering Physics Institute, Moscow, 115409 Russia}
\begin{abstract}
A tight-binding model of 8-$Pmmn$ borophene, a two-dimensional boron crystal, is developed. We confirm that the crystal hosts massless Dirac fermions and the Dirac points are protected by symmetry.
Strain is introduced into the model, and it is shown to induce a pseudomagnetic field vector potential and a scalar potential. The dependence of the potentials on the strain tensor is calculated.
The physical effects controlled by pseudomagnetic field are discussed.
\end{abstract}

\maketitle

\section{Introduction}

The recent research in nanoelectronics pay much attention to the two-dimensional crystals and heterostructures \cite{VdWreview}. The boron studies contributed to the advances in that field by successful synthesizing three different two-dimensional crystalline boron structures \cite{bbSynth15,bbSynth16}. Two-dimensional boron crystals are referred to as borophenes \cite{bbNatComm14,bbSynth16,bbSynth15}.

Theoretical predictions of borophene structures and their properties are being developed intensively, see \cite{bbNatComm14,bbSynth15,bbOganov,boron2Drev,advXrev} and references therein. One of the most stable predicted structures, $Pmmn$ boron, is shown to be a~Dirac semimetal \cite{bbOganov}. That material was studied in detail in Ref.~\cite{bb8Pmmn}, where the term ``8-$Pmmn$ borophene'' was introduced, which we will use hereafter. Dirac semimetals have zero energy gap and conical dispersion law of low-energy electronic excitations. The physics of the 2D and 3D Dirac semimetals and related materials such as Weyl semimetals is also an actively developing field \cite{DSMAnnuRev,DSMJPCM,rare2Drev,advXrev}. The first 2D Dirac material discovered, graphene, remains to be a focus of intense research in connection with various applications including but not limited to nanoelectronics.
The study of Dirac semimetal $Pmmn$ borophene is interesting since it shares some properties of graphene but shows differences in other aspects, for example, the dispersion law of its low-energy excitations is anisotropic, in contrast to graphene \cite{bbOganov}.

Graphene shows numerous remarkable effects, one of which is the strain-induced elastic pseudomagnetic gauge field \cite{pmfreview,gpmfAndo,gpmfManies,GenHamGr}. Inhomogeneous strain induces an effective field which can be as strong as tens and hundreds of teslas, which has been confirmed experimentally \cite{pmfexp,Yeh}. The particles in graphene feel that effective field just in the same way as an external magnetic field except that the strain-induced field does not break the time-reversal symmetry. It is reasonable to expect the same effect in other Dirac materials. This effect was shown to be present in a simple 3D Dirac semimetal model \cite{dsmpmf}.

In the present work we show that the Dirac semimetal 8-$Pmmn$ borophene structure under strain indeed exhibits giant pseudomagnetic field as well as scalar potential. We develop a~tight-binding model to describe electronic structure of borophene. We analyze the symmetries of the system to find what contributions to the low-energy Hamiltonian from small, moderately non-uniform strain are allowed. We present numerical results for the vector and scalar potentials arising in the strained lattice and discuss possible experimental probing of the effect.

The paper is organized as follows: in section~\ref{sec:tb} we introduce the tight-binding Hamiltonian of 8-$Pmmn$ borophene, discuss its main features and give the values of its parameters obtained from fitting; in section~\ref{sec:pmf} we include strain using the symmetries and the numerical treatment and discuss possible experimentally detectible effects; we conclude in section~\ref{sec:concl}.

\section{Tight-binding model \\ of 8-$Pmmn$ borophene}\label{sec:tb}

\begin{figure}[t]
\centering
\includegraphics[width=\columnwidth]{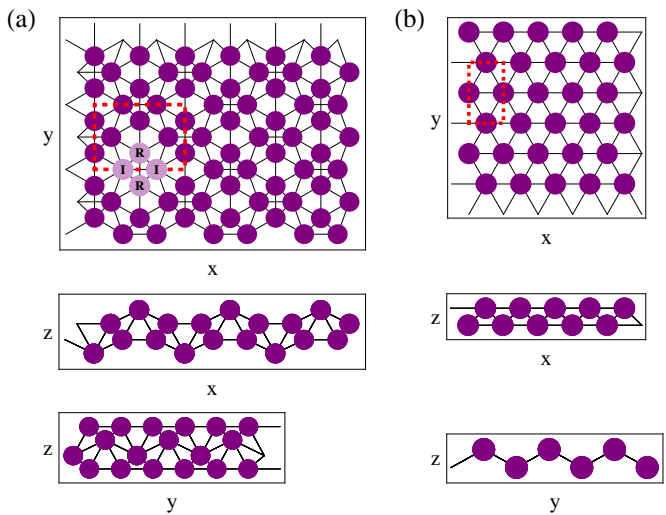}
\caption{\label{fig:struct}(Color online) Crystal structure of $Pmmn$ boro\-phenes. Dashed line encloses the unit cell.
(a)~Crystal structure of the Dirac semimetal 8-$Pmmn$ borophene predicted in \cite{bbOganov}. I and R labels mark the inner and ridge atoms \cite{bb8Pmmn}.
(b) Crystal structure of the metallic 2-$Pmmn$ borophene predicted in \cite{kuqu} and synthesized in \cite{bbSynth15}.}
\end{figure}

Two of the possible borophene structures, the predicted one from Ref.~\cite{bbOganov} shown in Fig.~\ref{fig:struct}(a) and the experimentally obtained one from Ref.~\cite{bbSynth15} (which had been first predicted in Ref.~\cite{kuqu}) shown in Fig.~\ref{fig:struct}(b) are both two-dimensional but not flat, and they share the same space group $Pmmn$ (No. 59 in \cite{itcbook}).
Since these structures have 8 and 2 atoms per unit cell respectively, we will call them 8-$Pmmn$ borophene (following \cite{bb8Pmmn}) and 2-$Pmmn$ borophene.
We are mostly interested in the former since it hosts Dirac fermions and thus can exhibit strain-induced pseudomagnetic field, but symmetry allows to make some conclusions regarding both those structures.

\begin{figure}[t]
\centering
\includegraphics[width=0.5\columnwidth]{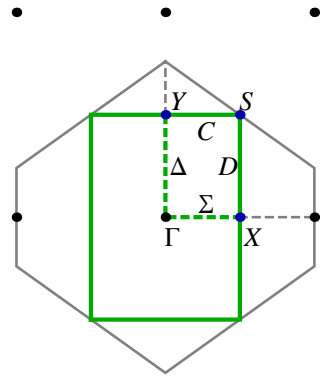}
\caption{\label{fig:bz}(Color online)
Reciprocal space of a 2D crystal with $Pmmn$
space group: the black dots are the reciprocal lattice points, the thick solid green line encloses the rectangular first Brillouin zone, the blue dots are the high-symmetry points of the Brullouin zone, the dotted lines are the internal high-symmetry lines. Hereinafter we use the Bilbao Crystallographic Server \cite{\Bilbao} notation which differs from the one used in \cite{bbOganov} by the substitution $X\leftrightarrow Y$. The thin gray line encloses the hexagonal extended Brillouin zone of the $n$-symmetrized wave functions which are described by $Cmmm$ group.
}
\end{figure}

The $Pmmn$ space group is (redundantly) generated by the translations by the vectors $(a,0,0)$, $(0,b,0)$, and $(0,0,c)$ (the latter is lacking for the two-dimensional case), inversion $I$, rotation $C_2$, reflection $\sigma_\mathrm{v}$,
and the $n$-glide plane consisting of the $(a/2,b/2,0)$ translation and the $z\to-z$ reflection. We also neglect the possible spin-orbit coupling, so the system has the time-reversal symmetry $\theta$. The Hilbert space is then split into two subspaces of the $n$-symmetric and the $n$-antisymmetric wave functions, and the Hamiltonian $H(k_x,k_y)$ can be brought to the block-diagonal form with the blocks $H_\mathrm{S}$ and $H_\mathrm{A}$. The Bloch waves from every subspace then correspond to the smaller effective unit cell (Fig.~\ref{fig:struct}) and doubled first Brillouin zone (Fig.~\ref{fig:bz}).
It is well known that glide-symmetric crystals have Hamiltonian consisting of such blocks (see, e. g., \cite{symmDirac2D,symmNatPhys}), these blocks being related by $H_\mathrm{A}(k_x,k_y) = H_\mathrm{S}(k_x+2\pi/a,k_y) = H_\mathrm{S}(k_x,k_y+2\pi/b)$. For a tight-binding model of an arbitrary crystal with $Pmmn$ symmetry, that is shown directly in the Appendix. Then it follows that there is a twofold glide-symmetry-protected degeneracy at the boundary of the original Brillouin zone (lines $C$ and $D$ in Fig.~\ref{fig:bz}).
As a consequence, if the number of boron atoms in the unit cell is not a multiple of four, as in the case of 2-$Pmmn$ borophene, and we do not have even number of electrons per unit cell (leaving aside twofold spin degeneracy) at the charge neutrality, then the crystal must be metallic. This applies to the material described in Refs.~\cite{kuqu,bbSynth15}.

The two inequivalent atoms in the lattice of 8-$Pmmmn$ borophene are inner atoms and ridge atoms, using the terminology of Ref.~\cite{bb8Pmmn}.
The lattice is characterized by the following parameters \cite{bbOganov}: $a=0.452$ nm, $b=0.326$ nm, the coordinates of two inequivalent atoms are $(a/2, 0.247b, 0.109\textrm{\ nm})$ (ridge atom) and $(0.185a, b/2, 0.040\textrm{\ nm})$ (inner atom), and the coordinates of other atoms are obtained via symmetry operations.

To investigate the electronic structure of the 8-$Pmmn$ borophene further we develop a tight-binding model. For each of the 8 atoms of the unit cell we take into account one $2s$ orbital and three $2p$ orbitals, ignoring spin. In the basis of symmetric/antisymmetric combinations of $s,p_x,p_y$ orbitals of the atoms related by $n$ symmetry and antisymmetric/symmetric combinations of their $p_z$ orbitals the Hamiltonian takes the block-diagonal form with two blocks, $H_\mathrm{S}$ and $H_\mathrm{A}=H_\mathrm{S}(k_x+2\pi/a,k_y)$, which define equivalent 16-band models. Any of them can be taken to study low-energy dynamics.

For the Hamiltonian two-center integrals $V_i(R)$, $i=ss\sigma,sp\sigma,pp\sigma,pp\pi$, we use NRL parametrization \cite{bib-nrl}
\begin{equation}
V_i = (e_i+f_iR+g_iR^2)\exp(-h_i^2R).
\label{nrl}
\end{equation}
The hoppings are constructed in a usual way \cite{lcao}, orbital overlap is neglected. So we have sixteen parameters for the matrix elements. We take into account only the six inequivalent bonds with length less than 2~\AA, which are the ones shown in Fig.~\ref{fig:struct}(a). The on-site energies $\epsilon_s$ and $\epsilon_p$ are taken to be the same for all atoms, which adds one parameter $\epsilon_p-\epsilon_s$. To obtain those 17 parameters we fit the valence band and the conduction band to the GGA-PBE DFT data of \cite{bbOganov}. The parameters given in \cite{b-nrl} were used as an initial guess for the fitting. $\epsilon_s$ is extracted from the DFT total energy. The resulting parameters are given in Table~\ref{tab:params}.

\begin{table}[b]
\centering
\begin{tabular}{lcccc}
\hline
bond type & $ss\sigma$ & $sp\sigma$ & $pp\sigma$ & $pp\pi$ \\
\hline
$e$ [Ry]                    &$-45.61$&  7.926 &  0.770 & 0.987 \\
$f$ [Ry/$r_\mathrm{B}$]     &$ -1.30$&$-2.550$&  1.159 & 0.864 \\
$g$ [Ry/$r_\mathrm{B}^2$]   &   2.90 &$-7.757$&  2.286 & 0.889 \\
$h$ [$r_\mathrm{B}^{-1/2}$] &   5.28 &  0.937 &$-3.485$& 1.335 \\
\hline
$\epsilon_p-\epsilon_s$ [Ry] & 0.523 &&&\\
~~~~~~$\epsilon_s$~[Ry] & 0.057 &&&\\
\hline
\end{tabular}
\caption{\label{tab:params}
Fitting parameters in (\ref{nrl}). $r_\mathrm{B}=0.0529$ nm is the Bohr radius.}
\end{table}

\begin{figure}[t!]
\centering
\includegraphics[width=\columnwidth]{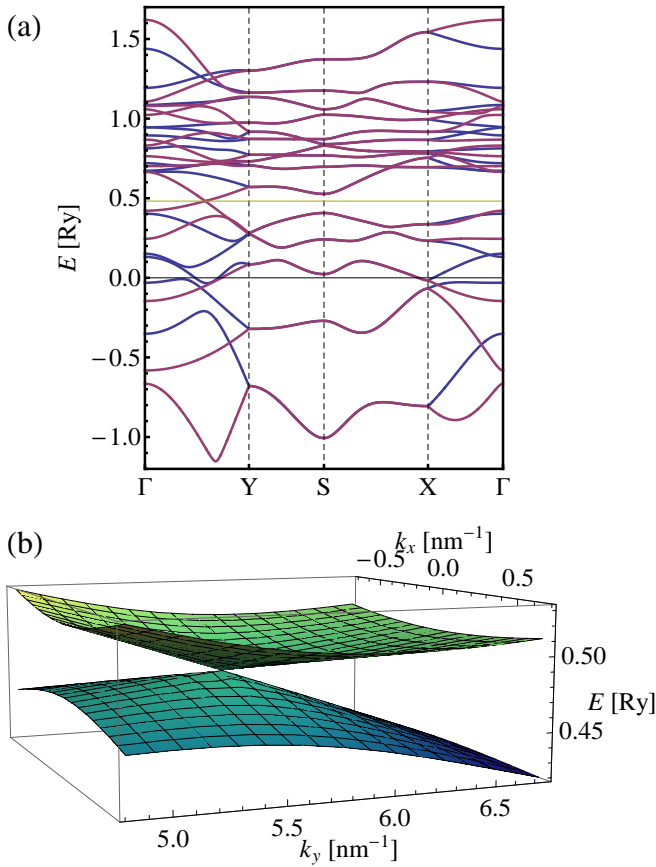}
\caption{\label{fig:bands}(Color online)
Band structure of 8-$Pmmn$ borophene.
(a) Band structure. The special points are explained in Fig.~\ref{fig:bz}. The Fermi level is shown with a yellow line. Blue and purple lines, respectively, correspond to bands defined by $H_\mathrm{S}$ and $H_\mathrm{A}$. One can see the twofold degeneracy at the boundary of the Brillouin zone and the Dirac point at the Fermi level. Two lowest energy bands are omitted.
(b) The tilted, anisotropic Dirac cone in the vicinity of point $\mathbf{k}_\mathrm{D}$.
}
\end{figure}

The resulting tight-binding model gives good agreement with the DFT data of \cite{bbOganov} and reproduces some features of the more detailed DFT data of \cite{bb8Pmmn}. The band structure is shown in Fig.~\ref{fig:bands}. The bands of $H_\mathrm{S}$ and $H_\mathrm{A}$ taken separately do not cross each other, there are only Dirac-point-touchings in the band structure of each block (however, when they are taken together, there are many band crossings, including the ones on the boundary of Brillouin zone). In particular, we obtain dispersion of energy excitations near the Fermi energy as tilted anisotropic Dirac cones at $\mathbf{k}_\mathrm{D}=(0,k_\mathrm{D})$ and $-\mathbf{k}_\mathrm{D}$, $k_\mathrm{D}=0.290\times2\pi/b$.

Regarding the charge density distribution of the low-energy Dirac electronic states in our model, the conduction band and the valence
band share the same main features in this aspect. The contributions from the two inequivalent types of atoms are virtually equal but originate from different
orbitals: the inner atoms mostly contribute to Dirac states by $p_z$ states, while in the case of the ridge atoms the contribution of the three $p$ orbitals are of
the same order ($s$ orbitals contribute much less). It seems established that the $p_z$ states of the inner atoms are vital for the Dirac states in
8-$Pmmn$ borophene. However, the role of the $p$ states localized on the ridge atoms is controversial. Indeed: the DFT results of Ref.~\cite{bbOganov} show that their $p_z$
orbitals contribute to the Dirac states, our tight-binding model gives equally important contribution of all three $p$ orbitals, and the DFT results of Ref.~\cite{bb8Pmmn} deny the contribution of the ridge atoms completely.

As pointed out in \cite{symmDirac2016}, the band touching at $\Delta$ line of the Brillouin zone (see Fig.~\ref{fig:bz}) is protected by the $\sigma_\mathrm{v}$ ($x\to-x$) reflection symmetry. The band touching, i. e. the Dirac point, arises as the crossing of two one-dimensional bands which belong to different subspaces of that symmetry. One of those bands corresponds to the wave functions symmetric with respect to $\sigma_\mathrm{v}$ reflection and the other one corresponds to the wave functions antisymmetric with respect to $\sigma_\mathrm{v}$.
So the Dirac point is protected by symmetry; however, the robust protection of the degeneracies, which is usually provided by the glide plane symmetries \cite{symmDirac2D}, is not present in this case.

\section{Strain-induced \\ pseudomagnetic field}\label{sec:pmf}
\subsection{Dirac Hamiltonian}

A general tilted, anisotropic two-dimensional Dirac cone shape is described by five parameters; $\sigma_\mathrm{v}$ symmetry reduces this number to three ($v_x, v_y, v_\mathrm{t}$):
\begin{equation}
E(\mathbf{p}) = p_yv_\mathrm{t}\pm\sqrt{p_x^2v_x^2+p_y^2v_y^2}.
\label{ddisp}
\end{equation}

The corresponding low-energy two-band effective massless Dirac Hamiltonian in the vicinity of $\mathbf{k}=\mathbf{k}_\mathrm{D}$ can be written as
\begin{equation}
H_\mathrm{D} = v_x\sigma_xp_x+v_y\sigma_yp_y+v_\mathrm{t}\sigma_0p_y
\label{dham}
\end{equation}
in a proper basis. The energy offset is the charge neutrality point of the $Pmmn$-borophene, i.~e., the Dirac point.

The velocities $v_x,v_y,v_\mathrm{t}$ are related to the original tight-binding Hamiltonian through its characteristic polynomial $P(E,\mathbf{k})$ or its eigenvalues $E(\mathbf{k})$ via the following formulas:
\begin{eqnarray}
v_\mathrm{t} = -\frac{\partial^2P/\partial k_y\partial E}{\partial^2P/\partial E^2},\nonumber\\
v_x^2 =-\frac{\partial^2P/\partial k_x^2}{\partial^2P/\partial E^2},\quad
v_y^2 = v_\mathrm{t}^2-\frac{\partial^2P/\partial k_y^2}{\partial^2P/\partial E^2};\\
\fpp{E}{k_x}=\pm v_x,\quad \fpp{E}{k_y}=\pm v_y+v_\mathrm{t}. \label{slopes}
\label{}
\end{eqnarray}
All the derivatives are taken at $\mathbf{k}=\mathbf{k}_\mathrm{D}$, $E=0$. The second Dirac point at $\mathbf{k}=-\mathbf{k}_\mathrm{D}$ has the opposite chirality, the opposite sign of $v_\mathrm{t}$; the signs of $v_x$ and $v_y$ depend on the basis.

Our model gives $v_x=0.86\times10^6$ m/s, $v_y=0.69\times10^6$ m/s, $v_\mathrm{t}=0.32\times10^6$ m/s, so the slopes of the dispersion in $y$ direction, $v_y\pm v_\mathrm{t}$ according to (\ref{slopes}), are $1.01\times10^6$ m/s and $0.37\times10^6$ m/s. The Dirac cone is shown in Fig.~\ref{fig:bands}(b).

\subsection{Symmetry-allowed terms in the Hamiltonian}\label{sec:symm}

Now we consider the strained lattice in the same way as it was done in \cite{GenHamGr} for the case of graphene. The strain is described by the displacement field $\mathbf{u}(\mathbf{r})$ and the strain tensor which equals $u_{ij}(\mathbf{r}) = \frac12\left(\fpp{u_i}{r_j}+\fpp{u_j}{r_i}\right)$ in the first order with respect to the derivatives of $u_i$. Small, moderately inhomogeneous strain can contribute to the Hamiltonian some terms which must be invariant under the symmetry operations which leave the crystal structure and also $\mathbf{k}_\mathrm{D}$ invariant \cite{bradleybook,bassanibook}. $n$-glide plane contributes nothing here, so we have two symmetry operations: the $\sigma_\mathrm{v}$ reflection ($x\to-x$) and the $\theta C_2$ operation ($\pi$ rotation together with time reversal: $x\to-x,y\to-y,t\to-t$). Action of each of these two operations upon various quantities such as components of the strain tensor or the momentum can be represented either as 1 or as $-1$. Since the time reversal $\theta$ is merely the complex conjugation, it changes the sign of $i$. Then the invariance of the Hamiltonian [both the full one and the low-energy one (\ref{dham})] and the equality $\sigma_z=-i\sigma_x\sigma_y$ give us the results summarized in Table~\ref{tab:symm}.

\begin{table}[t]
\centering
\begin{tabular}{ccc}
\hline
action of symmetries &~~~& quantities \\
\hline
$\sigma_\mathrm{v}=+1$, $\theta C_2=+1$ & & $\sigma_0, H, p_y, \sigma_y, u_{xx}, u_{yy}$ \\
$\sigma_\mathrm{v}=-1$, $\theta C_2=+1$ & & $p_x, \sigma_x, u_{xy}, \omega$ \\
$\sigma_\mathrm{v}=+1$, $\theta C_2=-1$ & & $i, y, \fpp{}{y}, u_y$ \\
$\sigma_\mathrm{v}=-1$, $\theta C_2=-1$ & & $\sigma_z, x, \fpp{}{x}, u_x$ \\
\hline
\end{tabular}
\caption{\label{tab:symm}
The action of symmetry operations upon various quantities. The Pauli matrices $\sigma_{0,x,y,z}$ are understood as operators acting on wave functions written in the basis in which the low-energy Hamiltonian has the form (\ref{dham}).
}
\end{table}

As can be seen from Table~\ref{tab:symm}, the following invariant quantities involving the strain tensor are possible: $u_{xx}\sigma_0$ and $u_{yy}\sigma_0$ (scalar potential -- an energy shift), $u_{xx}\sigma_y$ and $u_{yy}\sigma_y$ (shift of $p_y$), $u_{xy}\sigma_x$ (shift of $p_x$), and the higher-order terms which are small for small, smooth strains. The coefficients of these terms depend on the Hamiltonian parameters and must be obtained from the model. The shift of the momentum components may be interpreted as the vector potential of the strain-induced pseudomagnetic field.

Table~\ref{tab:symm} also mentions the local rotation $\omega=\frac12\left(\fpp{u_x}{y}-\fpp{u_y}{x}\right)$. Even though this quantity itself cannot give any observable contribution, its derivatives can enter the Hamiltonian. For example, the term $\sigma_z\fpp{\omega}{y}$ is invariant under the $\sigma_\mathrm{v}$ and $\theta C_2$ symmetries. However, this term is small for smooth strains.

The gap could be opened by a term proportional to $\sigma_z$. It can be seen from Table~\ref{tab:symm} that strains cannot induce such a term in the lowest order. The above-given example $\sigma_z\fpp{\omega}{y}$ shows that higher-order terms can have such form, so in principle a very small gap can be opened by strains. However, it seems that in graphene, where it is also possible \cite{GenHamGr}, such mechanism of gap opening was not observed for now.

So the symmetry forbids gap opening for small, moderately inhomogeneous strains and allows shifting of the Dirac point, thus allowing pseudomagnetic and pseudo-electric fields of the certain forms.

\subsection{Numerical results for the potentials}

We performed the numerical investigation of the tight-binding Hamiltonian introduced in Sec.~\ref{sec:tb}. The strains were taken into account via the change in matrix elements, which is contributed by the change of the inter-atomic distances [changing hoppings in accordance with (\ref{nrl})] and the change of the directions of inter-atomic vectors changing the coefficients of hoppings in accordance with \cite{lcao}.

We indeed obtained the potentials consistent with the results of Sec.~\ref{sec:symm}. As a result, the effective Hamiltonian of deformed material becomes:
\begin{multline}
H=v_x\sigma_x\left(p_x-\tfrac{e}{c}A_x(u_{ij})\right)+\\
+(v_y\sigma_y+v_\mathrm{t}\sigma_0)\left(p_y-\tfrac{e}{c}A_y(u_{ij})\right) + e\varphi(u_{ij}).
\label{}
\end{multline}
\begin{eqnarray}
\varphi = V_{xx}u_{xx}+V_{yy}u_{yy}; \label{bpef}\\
\mathbf{A} = (\alpha_{xy}u_{xy},\ \alpha_{xx}u_{xx} + \alpha_{yy}u_{yy}). \label{bpmf}
\end{eqnarray}
\begin{eqnarray}
V_{xx}=-6.2\text{ V, }V_{yy}=-6.0\text{ V;} \label{numphi}\\
\alpha_{xy}=3.63\text{ G$\times$cm,} \label{numa}\nonumber\\
\alpha_{xx}=3.58\text{ G$\times$cm, }\alpha_{yy}=-1.15\text{ G$\times$cm.}
\end{eqnarray}
The scalar potential $\varphi$ has the same sign in the $-\mathbf{k}_\mathrm{D}$ valley, while the vector potential $\mathbf{A}$ has the opposite sign there. The effective fields are given by $\mathbf{E}=-\nabla\varphi$, $\mathbf{B}=\nabla\times\mathbf{A}=(0,0,\fpp{A_y}{x}-\fpp{A_x}{y})$. Note that equations (\ref{bpef}--\ref{bpmf}) differ from the case of graphene which has different symmetry and $\varphi=V(u_{xx}+u_{uu})$, $\mathbf{A}=\alpha(u_{xx}-u_{yy}, -2u_{xy})$.

The figures in Eq. (\ref{numphi}) are missing the contribution to the scalar potential from the change of the on-site potentials in a strained crystal, but general Eq. (\ref{bpef}) for the scalar potential is true. It leads to the static charging effect, however, it can be suppressed by the substrate \cite{Yeh}.

On the other hand, the figures in Eqs.~(\ref{numphi}) and (\ref{numa}) must be noticeably overestimated because of the straightforward treatment of strain.
The strain field actually provides the description for the deformation of the Bravais lattice, and in complex lattices such as that of 8-$Pmmn$ borophene the atoms
inside a unit cell tend to relax, and so the strain-induced potentials get renormalized and thus decreased \cite{Midtvedt}.

\subsection{Effects}

From (\ref{bpmf}), (\ref{numa}) it can be seen that if the characteristic size of strain inhomogeneity is about tens of nanometers then the pseudomagnetic field can be as large as millions of gausses (i. e., hundreds of teslas) as in graphene. So it is possible to engineer giant effective pseudomagnetic field via strain to control electric currents in borophene. In particular, since the pseudomagnetic field has different signs for two valleys, that controllable field may be used in valleytronics based on Dirac materials \cite{vtronics10,vtronics13}.

\begin{figure}[t]
\centering
\includegraphics[width=0.8\columnwidth]{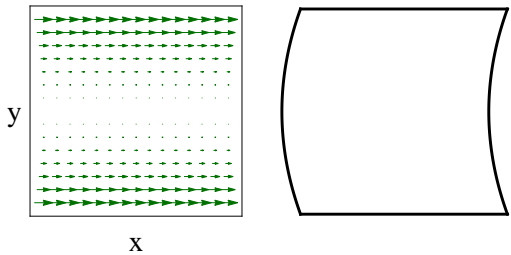}
\caption{\label{fig:strain}
Strain field $\mathbf{u}(\mathbf{r})=(y^2/b,0,0)$ in 8-$Pmmn$ borophene which provides a homogeneous pseudomagnetic field with no scalar potential, and the shape of a deformed borophene flake.
}
\end{figure}

Certain strains such as the one shown in Fig.~\ref{fig:strain} provide homogeneous pseudomagnetic field. In the paper \cite{bedt},
the Landau levels in the system with anisotropic, tilted massless Dirac dispersion were studied, and it was found that for large $N$, the energy of the $N$-th Landau level in strong magnetic field $B$ is approximately proportional to $\sqrt{|BN|}$ as in the case of usual Dirac cone.

Another phenomenon that could be observed in strained 8-$Pmmn$ borophene is the quantum Hall effect. It would require crossed pseudomagnetic and electric (or pseudo-electric) fields, so because of valley-dependent sign of the pseudomagnetic field and the absence of the external magnetic field we would obtain anomalous quantum valley Hall effect \cite{vtronics10,qhall}.
Besides, the interplay between real and pseudo-magnetic fields in a Dirac material could lead to odd integer quantum Hall effect \cite{oddiqhe}.

Besides, pseudomagnetic field in a strained Dirac material provides significant Faraday and Kerr rotation when external field is also present \cite{faraday,faraday16}.

The main feature of borophene compared to graphene is its anisotropic Dirac cone which originates from different lattice symmetry.
For the strain-induced potentials, it manifests as follows.
Say, we have an arbitrary strain field and rotate it as a whole. The strain tensor then transforms $u_{ij}(\mathbf{r}) \to u_{ij}'(\mathbf{r}')$ in the usual way,
where $\mathbf{r}$ is related to $\mathbf{r}'$ via some orthogonal transformation. Will the strain-induced potentials stay the same?
For the case of graphene, the scalar potential $\varphi=(u_{xx}+u_{yy})V$ and the norm of the vector potential $\mathbf{A}=(u_{xx}-u_{yy},-2u_{xy})A$ \cite{pmfreview}
would both stay the same: $\varphi(u'_{ij}(\mathbf{r}')) = \varphi(u_{ij}(\mathbf{r}))$, $|\mathbf{A}(u'_{ij}(\mathbf{r}'))|^2 = |\mathbf{A}(u_{ij}(\mathbf{r}))|^2$.
However, for the case of borophene that is generally not true.
Thus it is possible in principle to deform a graphene or 8-$Pmmn$ borophene flake in some way,
then deform it again with the same deformation pattern rotated, and compare the static charging patterns.
They would be the same for the case of graphene but different in the case of borophene.

In view of the above we believe that it will be interesting to synthesize 8-$Pmmn$-borophene for these experiments to become possible. Since it is predicted to be one of the most stable borophene structures, its synthesis via molecular beam epitaxy or some of the chemical techniques (which are used to obtain 3D boron allotropes \cite{boronbook}) seems possible.

\section{Conclusion}\label{sec:concl}

We have developed a tight-binding model of a 2D Dirac semimetal 8-$Pmmn$ borophene predicted in \cite{bbOganov}.
The parameters were obtained from a fit to the DFT data and the resulting electronic structure reproduces the main features of the DFT band structure, including the Dirac points at the Fermi level.
Symmetry analysis have shown that the Dirac points are symmetry-protected.

We have analyzed the general form of contribution of strains to Hamiltonian using the symmetry considerations.
Using the tight-binding model we have obtained strain-induced scalar and vector potential in 8-$Pmmn$ borophene expressed through the strain tensor, equations (\ref{bpef}--\ref{numa}).

The pseudomagnetic field can be as large as hundreds of teslas. It should be detectable in 8-$Pmmn$ borophene through the Landau quantization, Faraday effect, and quantum valley Hall effect, and it can be applied to valleytronics device development.

\begin{acknowledgments}
This work was supported by Russian Foundation for Basic Research grant No.~14-02-01059.
\end{acknowledgments}

\appendix*\section{Hamiltonian of a system with $n$-glide symmetry}\label{sec:nhamsymm}
Let the tight-binding Hamiltonian $H_\mathrm{tb}$ include $2N$ orbitals per unit cell, $N=4\times4=16$ for our 8-$Pmmn$ borophene model. In the basis of these atomic orbital wave-functions, if we replace $\mathbf{k}\to\mathbf{k}+\Delta\mathbf{k}$ with $\Delta\mathbf{k}$ being a reciprocal lattice vector then this is equivalent to the unitary transformation of the Hamiltonian $U=\diag\{e^{-i\Delta\mathbf{k}\mathbf{r}_i}\}_{i=1}^{2N}$ (a diagonal matrix with the elements $e^{-i\Delta\mathbf{k}\mathbf{r}_i}$ in its main diagonal) where $\mathbf{r}_i$ is the position of the atom on which the $i$-th orbital is localized in the unit cell. We enumerate the orbitals in such a way that for $i\leq N$, the atom on which orbital $i+N$ is localized is related to the atom on which orbital $i$ is localized by the non-symmorphic $n$-glide symmetry. $e^{-i\Delta\mathbf{k}\mathbf{r}_i}=-e^{-i\Delta\mathbf{k}\mathbf{r}_{i+N}}$ with $\Delta\mathbf{k}=(2\pi/a,0,0)$ or $\Delta\mathbf{k}=(0,2\pi/b,0)$ since in both cases $\Delta\mathbf{k}(\mathbf{r}_{i+N}-\mathbf{r}_i)=\pi$, so $U$ can be written in the block-diagonal form as $U=\diag(U_N,-U_N)$ (a block-diagonal matrix with the blocks $U_N$ and $-U_N$ in its main diagonal, with $U_N$ also diagonal) in that case.

To bring the Hamiltonian to the block-diagonal form, we perform the unitary transformation $S$ which is written as follows: for $i\leq N$, the matrix elements are given by $S_{ij}=(\delta_{ij}\pm\delta_{i+N,j})/\sqrt2$ where $\pm$ is plus or minus depending on how $\sigma_\mathrm{h}$ reflection acts on the type of orbital $i$ (in particular, it is plus for $s,p_x,p_y$ orbitals and minus for $p_z$ orbitals); for $i>N$, $S_{ij}=(\delta_{i-N,j}\mp\delta_{ij})/\sqrt2$. Then transformation $S$ brings the Hamiltonian to the block-diagonal form $SH_\mathrm{tb}S^\dagger=\diag(H_\mathrm{S},H_\mathrm{A})$.

Now we note that if $U=\diag(U_N,-U_N)$ then $SU=S'U'=U'S'$ where $U'=\diag(U_N,U_N)$ and $S'$ is $S$ with top $N$ rows swapped with bottom $N$ rows, so $S'H_\mathrm{tb}{S'}^\dagger=\diag(H_\mathrm{A},H_\mathrm{S})$. Thus $H_\mathrm{S}(k_x+2\pi/a,k_y,k_z)$ and $H_\mathrm{S}(k_x,k_y+2\pi/b,k_z)$ are both related to $H_\mathrm{A}(k_x,k_y,k_z)$ by unitary transformation $U'$. This means that we have to find the eigenvalues for just any one of these two blocks since the full spectrum can be obtained directly from them. However, the Brillouin zone of each block $H_\mathrm{S,A}$ is doubled (Fig.~\ref{fig:bz}).

A corollary is that on the boundary of the original Brillouin zone (or only its side faces for the 3D case) the two blocks of the Hamiltonian are equivalent so there is an $n$-glide-symmetry-protected degeneracy.
So every $(2k-1)$-th band always crosses $2k$-th band which means that at odd filling factor the Fermi level can never lie in a gap
(or even at point band touching) since the two bands that would be separated by that gap actually touch each other throughout the boundary of the Brillouin zone.
Thus such material is a metal.
That is the case of the 2-$Pmmn$ borophene structure shown in Fig.~\ref{fig:struct}(b) at charge neutrality.
An~insulator or a~semimetal must have even filling factor which is the case of Fig.~\ref{fig:struct}(a).

\bibliography{borbib}

\end{document}